# New single-molecule magnet based on $Mn_{12}$ oxocarboxylate clusters with mixed carboxylate ligands, $[Mn_{12}O_{12}(CN\text{-}o\text{-}C_6H_4CO_2)_{12}(CH_3CO_2)_4(H_2O)_4]\cdot 8CH_2Cl_2$: synthesis, crystal and electronic structure, magnetic properties

Lyudmila A. Kushch[a*], Valentina D. Sasnovskaya[a], Alexey I. Dmitriev[a], Eduard B. Yagubskii[a*], Oksana V. Koplak[b], Leokadiya V. Zorina[c], Danil W. Boukhvalov[d]

[a]*Institute of Problems of Chemical Physics Russian Academy of Science, Semenov's av., 1, Chernogolovka 142432, MD, Russian Federation.*
[b]*Educational and Scientific Center of National Academy of Sciences and T. Shevchenko National University, Kiev 01033, Ukraine*
[d]*Institute of Solid State Physics Russian Academy of Sciences, Academician Ossipyan str., 2, Chernogolovka 142432, MD, Russian Federation*
[c]*School of Computational Sciences, Korea Institute for Advanced Study, Seoul 130-722, Korea*

*A new high symmetry $Mn_{12}$ oxocarboxylate cluster $[Mn_{12}O_{12}(CN\text{-}o\text{-}C_6H_4CO_2)_{12}(CH_3CO_2)_4(H_2O)_4]\cdot 8CH_2Cl_2$ (**1**) with mixed carboxylate ligands is reported. It was synthesized by the standard carboxylate substitution method. **1** crystallizes in the tetragonal space group $I4_1/a$. Complex **1** contains a $[Mn_{12}O_{12}]$ core with eight $CN\text{-}o\text{-}C_6H_4CO_2$ ligands in the axial positions, four $CH_3CO_2$ and four $CN\text{-}o\text{-}C_6H_4CO_2$ those in equatorial positions. Four $H_2O$ molecules are bonded to four Mn atoms in an alternating up, down, up, down arrangement indicating 1:1:1:1 isomer. The $Mn_{12}$ molecules in **1** are self-assembled by complementary hydrogen C-H···N bonds formed with participation of the axial o-cyanobenzoate ligands of the adjacent $Mn_{12}$ clusters. The lattice solvent molecules ($CH_2Cl_2$) are weakly interacted with $Mn_{12}$ units that results in solvent loss immediately after removal of the crystals from the mother liquor. The electronic structure and the intramolecular exchange parameters have been calculated. Mn 3d bands of **1** are rather broad and center of gravity of the bands shifts down from Fermi level. The overlap between Mn 3d bands and 2p ones of the oxygen atoms from carboxylate bridges is higher than in the parent $Mn_{12}$-acetate cluster. These changes in electronic structure provide significant difference in the exchange interactions in comparison to $Mn_{12}$-acetate. The magnetic properties have been studied on a dried (solvent-free) polycrystalline sample of **1**. The dc magnetic susceptibility measurements in the 2-300 K temperature range support a high-spin ground state ($S = 10$). A bifurcation*

*of temperature dependencies of magnetization taken under the zero field cooled and field cooled conditions observed below 4.5 K is due to slow magnetization relaxation. Magnetization versus applied dc field exhibited a stepwise hysteresis loop at 2 K. The ac magnetic susceptibility data revealed the frequency dependent out-of-phase ($\chi_M''$) signals characteristic of single-molecule magnets.*

**Introduction**

Over the past two decades, great interest has been shown in high-spin metal clusters, which reveal unusual mesoscopic magnetic properties on a scale of one molecule (superparamagnetisim, strong magnetic anisotropy, slow magnetic relaxation, blocking and quantum tunneling magnetization).[1,2] Such molecules with large spin in the ground state and negative magnetoanisotropy were named single-molecule magnets (SMMs). A large number of transition (3d, 4d, 5d), rare earth (4f) and actinide (5f) metal complexes with various nuclearities and topologies possessing the properties of SMMs is known today.[2-4] However, the most numerous and widely investigated class of SMMs is currently the family of [$Mn_{12}O_{12}(RCO_2)_{16}(H_2O)_4$] (R = $CH_3$, $CH_2Cl$, $CHCl_2$, $CF_2Cl$, $CF_3$, $C_6H_5$, $C_6F_5$, et al.) oxocarboxylate clusters. Their study has provided the discovery of molecular nanomagnetism[5] and gave the majority of current knowledge on this interesting magnetic phenomenon.[6] The crystal structures of these clusters contain a central cubane fragment $Mn^{IV}_4O_4$ surrounded by a ring of eight $Mn^{III}$ centers connected through bridging oxo ligands. Bridging carboxylate and terminal water ligands passivate the surface, so that each Mn center has an approximate octahedral coordination environment. Antiferromagnetic exchange coupling of four $Mn^{IV}$ ions with S = 3/2 and eight $Mn^{III}$ ions with S = 2 leads to the formation of the ferrimagnetic structure with total spin S = 10 in the ground state. A deep and systematic study of a $Mn_{12}$ family has shown that variation of R in the ligands does not significantly alter the geometry of the inner core of the molecule: the distances Mn-Mn and Mn-O-Mn and the angles Mn-O-Mn change by less than 2%. However, the change of the ligand may change the space group symmetry affecting both the arrangement of the molecules in a crystal and internal symmetry of the molecule. It was shown that such properties as size and electronegativity of the substitutes in the ligands influence strongly the degree of p-d hybridization, therefore the electronic structure of the SMMs.[7,8] Furthermore, incorporation of the additional functional groups into carboxylate ligands can lead to formation of $Mn_{12}O_{12}CR$ assemblies through various kinds of molecule interactions.[9] The supramolecular SMM dimmer [$Mn_4O_3Cl_4(EtCO_2)_3(py)_3$]$_2$ held together by close intermolecular Cl⋯Cl contact and C-H⋯Cl hydrogen bonds shows quantum magnetic behavior different from that of the discrete SMMs.[10] The weak coupling of two or more SMMs to each other is essential for SMM applications as qubits for quantum computation and as components in molecular spintronics devices, which would exploit their quantum properties (quantum tunneling of magnetization and quantum phase interference).[11] One would expect the presence of a strong electron-withdrawing CN-substituent in benzenecarboxylate ligand will lead to the clusters assembling due to formation of C-H⋯N≡C hydrogen bonds and/or C≡N⋯N≡C intermolecular contacts. Recently, we synthesized $Mn_{12}$ cluster with $p$-cyanobenzenecarboxylate ligand, [$Mn_{12}O_{12}(CN$-$p$-$C_6H_4CO_2)_{16}(H_2O)_4$], which unlike the known $Mn_{12}$ clusters is not dissolved in organic solvents providing an indication for strong intermolecular interactions which affect the magnetic behaviour.[9c] Here we studied the replacement of the bridging $CH_3CO_2^-$ ligands in the $Mn_{12}$-acetate cluster by $o$-cyanobenzoate bridges and prepared the new $Mn_{12}$ cluster with mixed carboxylate ligands, [$Mn_{12}O_{12}(CN$-$o$-$C_6H_4CO_2)_{12}(CH_3CO_2)_4(H_2O)_4$]·8$CH_2Cl_2$ (**1**). The synthesis, crystal and electronic structure as well as magnetic properties of **1** are presented.

**Experimental**

**General**

All preparations were performed under aerobic conditions. All chemicals and solvents were used as received. $Mn_{12}$-ac was prepared according to the procedure described in literature.[12a]

**Synthesis**

**[$Mn_{12}O_{12}(CN$-$o$-$C_6H_4CO_2)_{12}(CH_3CO_2)_4(H_2O)_4$]·8$CH_2Cl_2$ (1)**

A slurry of [$Mn_{12}O_{12}(CH_3CO_2)_{16}(H_2O)_4$]·2$CH_3COOH$·4$H_2O$ (0,5 g, 0.25 mmol) in $CH_2Cl_2$ (50ml) and ($C_2H_5$)$_2$O (15 ml) was treated with an excess of the CN-$o$-$C_6H_4COOH$ (1.18 g, 8 mmol). The mixture was stirred for 48 h. During this time all the solids were dissolved. Next 150 ml of hexane was added to the reaction solution, and the mixture was stored overnight in a refrigerator. The resulting solid was collected by filtration, and the above treatment was repeated three times. After four cycles of the treatment, 30 ml of diethyl ether was added to the resulting solid, and the mixture was stirred for 2.0 h to remove an excess of *o*-cyanobenzoic acid. The residue was filtered, washed with hexane and diethyl ether and dried in vacuum. The yield was 0.43 g (61%). Elemental analysis: Found: C, 42.88; H, 2.35; N, 5.77; O, 26.36; Mn, 22.63. Calc. for $C_{104}H_{68}Mn_{12}N_{12}O_{48}$ C, 42.87; H, 2,33; N, 5.77; O, 26.38; Mn, 22.64%. IR data $\nu_{max}$/cm$^{-1}$ 3450 (OH) from coordinated $H_2O$; 2240 (CN); 1610, 1590 (C=O from COO). The elemental analysis indicates a loss of lattice solvent (dichloromethane) upon drying relative to the crystallographically characterized species. The crystals for X-ray analysis were grown by layering a dichloromethane solution of the cluster with hexane. The crystals lose the solvent very easily and become unsuitable for crystallographic studies. To prevent the solvent loss, the crystals were kept in contact with the mother liquid.

**Physical measurements**

Analyses of C, H, N, O were carried out on a vario MICRO cube analyzing device. Infrared spectra (600 - 4000 cm$^{-1}$) were recorded using a Varian 3100 FTIR Excalibur Series spectrometer. Since the cluster crystals begin to lose the lattice solvent ($CH_2Cl_2$) immediately after their removal from the mother liquor, the magnetic properties have been studied on a dried (solvent-free) polycrystalline sample of **1**. Dc magnetic susceptibility studies were performed on a Quantum Design SQUID magnetometer equipped with a 5 T magnet and operating in the 2.0 – 300 K range. The experimental data were corrected for the sample holder and for the diamagnetic contribution calculated from Pascal constants. Ac magnetic susceptibility data were collected on the same instrument, employing a 4 G field oscillating at frequencies up to 1400 Hz. Magnetization vs field and temperature data were fit using the program ANISOFIT implemented on the MATLAB platform.[13]

**X-ray crystallography**

A suitable single crystal was placed into a glass capillary together with a drop of the mother liquid and transferred to the cold nitrogen stream of the Oxford Diffraction Gemini-R diffractometer. The melting temperature of both the solvents, dichloromethane and hexane, is *ca.* 180 K, and the sample was kept slightly above this temperature till full evaporation of the solvents followed by further cooling to 150 K. Full array of the X-ray data was collected at 150 K with Mo$K_\alpha$-radiation ($\lambda$ = 0.7107Å, graphite monochromator, ω-scan). Data reduction with empirical absorption correction of experimental intensities (Scale3AbsPack program) was made with the CrysAlisPro software.[14]

The structure was solved by a direct method followed by Fourier syntheses and refined by a full-matrix least-squares method using the SHELX-97 programs.[15] All non-hydrogen atoms except for solvent components were refined in an anisotropic approximation. The positions of H-atoms were calculated geometrically and refined in a riding model with isotropic displacement parameters $U_{iso}(H) = 1.2U_{eq}(C)$. Hydrogen atoms in terminal water ligands of the $Mn_{12}$ complex were localized in the difference electron density map and refined with restrained O-H bonds to be of equal length (SADI instruction), $U_{iso}(H) = 1.5U_{eq}(O)$.

**Crystal data.** $C_{112}H_{84}Cl_{16}Mn_{12}N_{12}O_{48}$, $M = 3592.39$, tetragonal, $a = 21.006(1)$, $c = 42.006(4)$ Å, $V = 18536(2)$ Å$^3$, $T = 150$ K, space group $I4_1/a$, $Z = 4$, $D_{calc} = 1.287$ g cm$^{-3}$, $\mu = 10.84$ cm$^{-1}$, 46525 reflections measured, 7694 unique ($R_{int} = 0.141$), 4072 reflections with $I > 2\sigma(I)$, 531 parameters refined, $R_1 = 0.112$, $wR_2 = 0.261$, GOF = 1.016. CCDC reference number is 885545.

**Electronic structure calculations**

To study the electronic structure and exchange interactions, we used a local density approximation (LDA) of density functional theory, taking into account the on-site Coulomb repulsion (LDA+U approach).[16] Accounting for the local Coulomb interaction is crucial for an adequate description of the transition metal oxide systems in general, and for $Mn_{12}$ SMMs in particular. The value of the Coulomb parameter U for $Mn_{12}$ clusters is 4 eV, as has been determined previously on the basis of experimental and theoretical studies.[7,8,17] For our electronic structure calculation, we used the LMTO-ASA[18] method implemented in the Stuttgart TB-47 code. Exchanges are calculated for Heisenberg spin-Hamiltonian $H=-J_{ij}S_iS_j$.

**Results and discussion**

**Synthesis**

The cluster $[Mn_{12}O_{12}(CN\text{-}o\text{-}C_6H_4CO_2)_{12}(CH_3CO_2)_4(H_2O)_4]\cdot 8CH_2Cl_2$ (**1**) was synthesized using known standard method by treating of $Mn_{12}$-acetate with an excess of $o$-cyanobenzoic acid in four cycles. The molar ratio of $Mn_{12}$-ac to ligands was 1 to 32 for all treatments. Repeating treatment is required because the ligand substitution is an equilibrium that must be driven to completion. However, unlike the reaction with *p*-cyanobenzoic acid[9c], a completely replaced product is not formed, despite four cycles of the treatment were performed.

$[Mn_{12}O_{12}(CH_3CO_2)_{16}(H_2O)_4]$ +16 CN-*o*-$C_6H_4CO_2H$ ↔

$[Mn_{12}O_{12}(CN$-*o*-$C_6H_4CO_2)_{12}(CH_3CO_2)_4(H_2O)_4]$ + 4 CN-*o*-$C_6H_4CO_2H$ + 12 $CH_3CO_2H$

We also studied the replacement of pivalate ligands of the $[Mn_{12}O_{12}((CH_3)_3CCO_2)_{16}(H_2O)_4]$ cluster by *o*-cyanobenzoate ones. It is known, that the $(CH_3)_3CCO_2$-ligands of $Mn_{12}$-pivalate are easily replaced by other carboxylates.[19] However, in the case *o*-cyanobenzoic acid the completely substituted product is not formed. One may speculate that this is due to the steric effect of *o*-cyanobenzoate ligand: the CN group locates adjacent to a carboxylate one.

**Crystal structure**

The $[Mn_{12}O_{12}(CN$-*o*-$C_6H_4CO_2)_{12}(CH_3CO_2)_4(H_2O)_4]·8CH_2Cl_2$ (**1**) complex crystallizes in the tetragonal space group $I4_1/a$ with four formula units per unit cell. The molecule of **1** is drawn in two projections in Fig. 1.

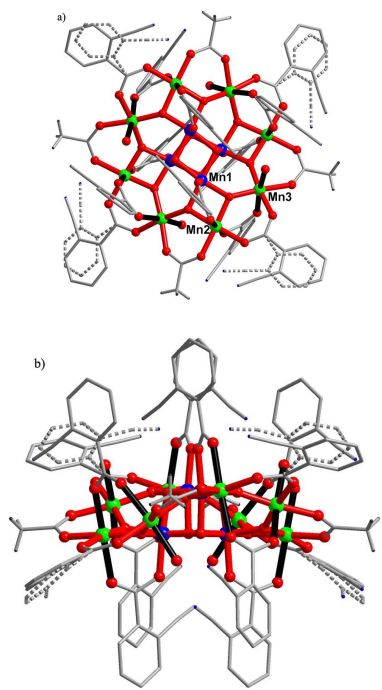

**Fig. 1** Structure of the complex **1** viewed along the crystal *c*-axis (a) and *b*-axis (b). $Mn^{IV}$ and $Mn^{III}$ ions are shown in blue and green colors, respectively. The thick black bonds indicate the Jahn-Teller elongation axes. H-atoms in CN-*o*-$C_6H_4CO_2$ ligands are omitted for clarity. Second position of the disordered ligand (30% occupancy) is shown by dashed lines.

The structure of the [Mn$_{12}$($\mu_3$-O$_{12}$)] core is similar to the previously characterized neutral [Mn$_{12}$] complexes.[6,12,20] There is a central [Mn$^{IV}_4$O$_4$] cubane moiety surrounded by a non-planar ring of eight Mn$^{III}$ ions which are bridged and connected to the cubane by eight $\mu_3$-O$^{2-}$ ions. The Mn$_{12}$ molecule is located on a four-fold inversion axis $\bar{4}$ and three of twelve manganese ions (Mn1-Mn3 in Fig. 1a) are crystallographically independent. Two outer Mn$^{III}$ ions differ by type of linking to the inner Mn$^{IV}$ ions: Mn2 is coordinated to a single Mn$^{IV}$ ion via two oxide bridges while Mn3 is coordinated to two Mn$^{IV}$ ions via two oxide bridges. Peripheral ligation of **1** is provided by twelve bridging CN-$o$-C$_6$H$_4$CO$_2^-$ ligands, eight of which are axial and four equatorial, four equatorial CH$_3$CO$_2^-$ ligands and four terminal water molecules in axial positions. The equatorial CN-$o$-C$_6$H$_4$CO$_2^-$ ligands are disordered between two slightly different orientations with 70 / 30% occupancy. Four H$_2$O molecules are bonded to four Mn3 atoms in an alternating up, down, up, down arrangement indicating 1:1:1:1 isomer which has been found previously in other neutral tetragonal Mn$_{12}$ complexes, including the parent Mn$_{12}$ acetate complex.[12, 20]

All the Mn centers are six-coordinated with near octahedral geometry. The Mn$^{IV}$ ions of the central cubane have close Mn-O bond lengths (1.847 – 1.923(6) Å) while high-spin Mn$^{III}$ ions of the outer ring show a Jahn-Teller distortion. Two *trans* Mn-O bonds of the Mn$^{III}$ octahedra of 2.113 – 2.189(6) Å are on the average 0.23 Å longer than the other four bonds lying in the range of 1.851 – 1.972(7) Å. The Jahn-Teller elongation axes, shown by thick black bonds in Fig. 1, are roughly normal to the disk-like [Mn$_{12}$O$_{12}$] core; corresponding angles to the crystal $c$-axis are 14.5 and 36.0º in Mn2 and Mn3 octahedra, respectively. This nearly parallel disposition of the Jahn-Teller axes makes the crystallographic $c$-direction the axis of easy magnetization. The angles between the Jahn-Teller axes of the adjacent Mn$^{III}$ octahedra are 24.9 and 30.3 °.

The Mn$_{12}$ molecules in **1** are self-assembled by complementary hydrogen C-H···N bonds formed with participation of the axial $o$-cyanobenzoate ligands of the adjacent Mn$_{12}$ clusters. There are two pairs of C-H···N bonds between every two clusters with H···N distances of 2.49 and 2.51 Å (C···N are 3.44(1) and 3.45(1) Å, angles C-H···N are 174.7 and 170.3º, respectively). Hydrogen bonding stabilizes a three-dimensional Mn$_{12}$ packing with near tetrahedral surrounding of every cluster by four nearest clusters (diamond-like packing, in contrast to body-centred packing with eight neighbours in most tetragonal Mn$_{12}$ structures). Presence of bulky CN-$o$-C$_6$H$_4$CO$_2$ ligands provides more porous structure of **1** in comparison with other $Mn_{12}$ complexes of tetragonal symmetry.[20] Large spaces between the $Mn_{12}$ units are filled by solvent molecules disordered throughout several sites of 20-60% occupancies. The analysis of additional solvent accessible voids in the structure with the PLATON program[21] shows four cavities in the unit cell, each being of 400 Å$^3$ in size and involving 90 electrons. The amount of electrons and the cavity volume correspond to two dichloromethane molecules per this void, therefore more accurate chemical formula of the compound should be apparently written as $[Mn_{12}O_{12}(CN-o-C_6H_4CO_2)_{12}(CH_3CO_2)_4(H_2O)_4]\cdot 10CH_2Cl_2$. However, these additional solvent molecules are not localized because of very strong positional disorder: different electron density peaks within the void do not exceed 1.5 ē/Å$^3$. The $CH_2Cl_2$ molecule showing the highest site occupation value (60%) forms the shortest C-H⋯O bond with $[Mn_{12}O_{12}]$ core: H⋯O and C⋯O distances are 2.39 and 3.36(1) Å, respectively; C-H⋯O angle is 164.1°. There are few other CH(solv.)⋯O,N($Mn_{12}$) and CH($Mn_{12}$)⋯Cl(solv.) contacts but all of them are not comparable with strong OH⋯O hydrogen bonds between the lattice $MeCO_2H$, $H_2O$ molecules and $[Mn_{12}O_{12}]$ core in the parent $Mn_{12}$ acetate complex.[12b]

Therefore, one may conclude that solvent molecules are weakly interacted with $Mn_{12}$ molecules in the complex **1** that results in solvent loss immediately after removal of the crystal from the mother liquor.

**Electronic structure**

To study electronic structure and exchange interactions in **1**, we performed the calculation using a local density approximation (LDA) of the density functional theory, taking into account the on-site Coulomb repulsion (LDA+U approach). The electronic structure for **1** (Fig.2) is closer to the experimentally and theoretically obtained electronic structure of $Mn_{12}O_{12}(C_6H_4SCH_3CO_2)_{16}(H_2O)_4$[17] with benzene rings in the ligands than to that of $Mn_{12}O_{12}(CH_3CO_2)_{16}(H_2O)_4$[7] with methyl groups in carboxylate fragments.

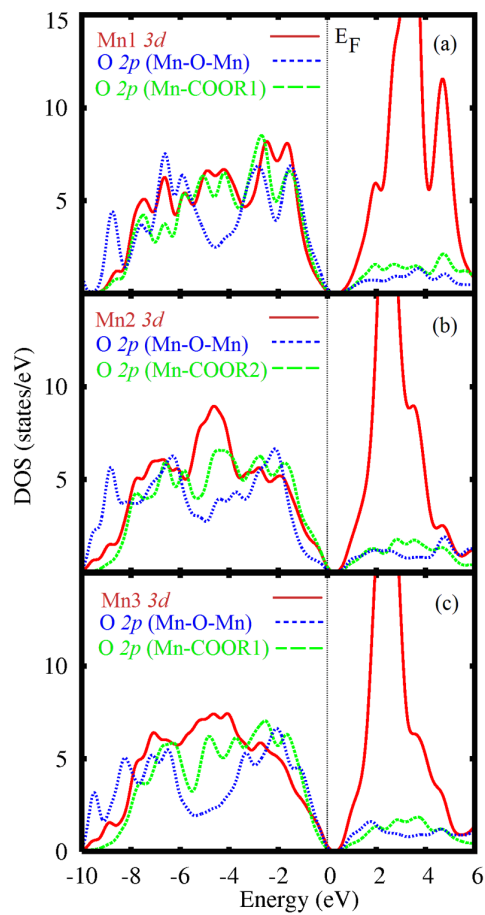

**Fig. 2** Partial densities of states for 3d orbitals of different manganese atoms (solid red lines), and oxygen 2p orbitals for two different types oxygen atoms – from Mn-O-Mn bridges (dotted blue lines) and carboxylate bridges with R1 = CN-o- $C_6H_4$- and R2 = $CH_3$ (dashed green lines).

Similarly to other $Mn_{12}$ SMMs with the aromatic rings in the ligands[8], 3d bands of manganese atoms of **1** are rather broad, and the center of gravity of the bands shifts down from Fermi level. The overlap between Mn 3d and O 2p bands is higher than in $Mn_{12}O_{12}(CH_3CO_2)_{16}(H_2O)_4$. These changes in the electronic structure provide rather robust charge transfer from ligands to $Mn_{12}O_{12}$ core and essential difference in the exchange interactions (Fig. 3) in comparison to starting $Mn_{12}O_{12}(CH_3CO_2)_{16}(H_2O)_4$. The evidence of significant charge transfer is the increase in magnetic moments of Mn(IV). These values are about 3.3$\mu_B$ for **1** in contrast to the values lower than 3.0$\mu_B$ for Mn12-ac.

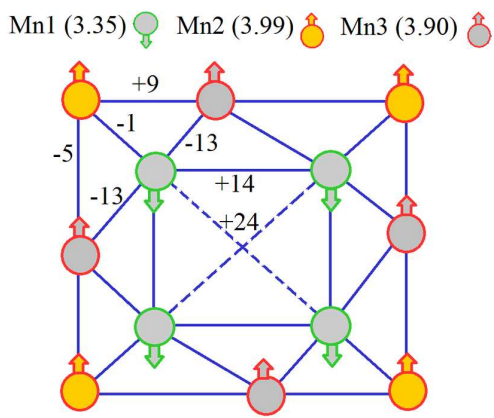

**Fig. 3** Calculated values of the magnetic moments (in $\mu_B$) and the exchange interactions (in K) for **1**.

The role of chemical composition of the ligands and the symmetry of the molecule in the exchange interactions of $Mn_{12}$ family have been discussed in Ref.[8] Similarly to other $Mn_{12}$ clusters with large ligands, the values of antiferromagnetic exchanges J(Mn1-Mn2) and J(Mn1-Mn3) decrease, and exchanges between Mn1 atoms in central cubane fragment turn to ferromagnetic and is rather large in **1**, Fig. 3. The obtained values of the exchange interactions are close to the values previously calculated for other members of $Mn_{12}$ family with high symmetry in the molecule and similar accumulated electronegativity of the ligands (25.95 for **1**).[8,17] The most significant issue of the magnetic interactions in **1** is the decay of J(Mn1-Mn2) to a negligible value.

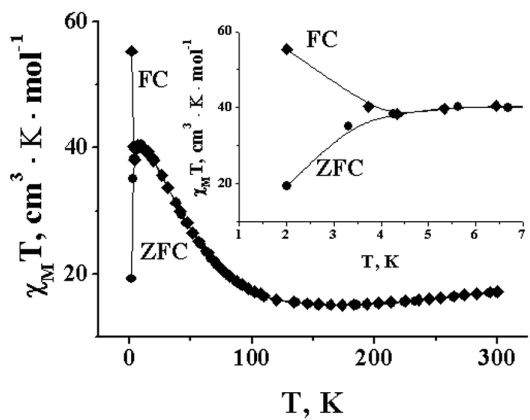

**Fig. 4** Temperature dependences of $\chi_M T$ for zero field cooled (ZFC) and field cooled (FC) experiments. The inset shows the behaviour of $\chi_M T$ (ZFC) and $\chi_M T$ (FC) in a low temperature range.

**Direct current (dc) magnetic susceptibility studies**

The variable-temperature dc magnetic susceptibility ($\chi_M$) data were collected on a dried polycrystalline sample of **1** in the 2.00-300 K range in a 1 T magnetic field in zero field cooled (ZFC) and field cooled (FC) regimes. The $\chi_M$T products for the ZFC and FC curves are shown in Fig. 4. At 300 K, the $\chi_M$T value is 17.1 cm³Kmol⁻¹, which is significantly lower than the theoretical value 31.5 cm³ K mol⁻¹ for magnetically non-interacting 8 Mn$^{III}$ (S=2) and 4 Mn$^{IV}$ (S=3/2) ions, indicating the presence of antiferromagnetic coupling within the cluster. On cooling below 300 K, the $\chi_M$T passes through a wide minimum at 170 K, and then increases rapidly to a maximum value of 40.5 cm³ K mol⁻¹ at 7.6 K, followed by a decrease down to 2 K (ZFC curve), Fig.4. The observed maximum for the $\chi_M$T is indicative of the stabilization of a high-spin ground state, whereas the sharp decrease of the $\chi_M$T at low temperatures can be attributed to zero-field splitting effect (ZFS). The behavior of $\chi_M$T vs T is typical for SMMs Mn$_{12}$ cluster family. A clear deviation of $\chi_M$T (FC) from $\chi_M$T (ZFC) is seen below 4.2 K (Fig. 4, the inset). Slow relaxation of the magnetization is probably responsible for the difference between ZFC and FC magnetization data. This fact indicates a relaxation process at 4.2 K.

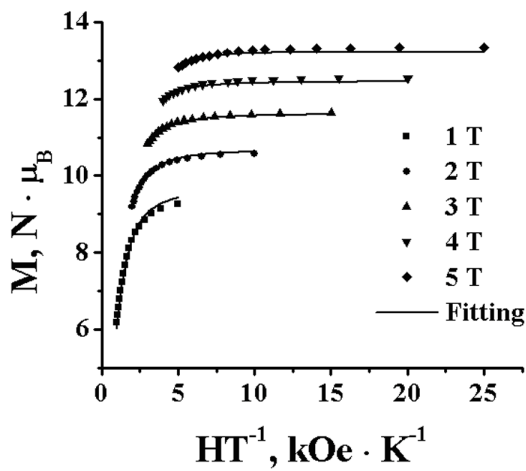

**Fig. 5** Isofield plots of magnetization $M/N\mu_B$ vs $H/T$ for a **dried** polycrystalline sample of **1** at indicated fields $H$. The solid lines show the fit to all of the data points, employing the method described in the text.

An examination of the ground spin state (S) and the magnitude of the axial zero-field splitting (D) for **1** was carried out by fitting the magnetization data using the program ANISOFIT.[13] The dc magnetization data were collected in the temperature range 2-10 K and at external fields of 1.0, 2.0, 3.0,

4.0, and 5.0 **T.** These data are depicted as M/Nμ$_B$ vs H/T (N is Avogadro's constant, μ$_B$ is the Bohr magneton) in Fig. 5. For a compound populating only the ground state and possessing no axial ZFS, the various isofield lines would be all superimposed, and M/Nμ$_B$ would saturate at value of gS. The nonsuperimposition of the isofield lines clearly indicates the presence of ZFS. The best fit for **1** was obtained with the parameters S = 10, g = 1.7 and $D = -0.39$ cm$^{-1}$ = $-0.56$ K. These values fall within the range that is typical for Mn$_{12}$ clusters, with the exception of the g tensor value which is rather low. Earlier such low g value was found for Mn$_{12}$O$_{12}$(C$_6$H$_5$C$_6$H$_4$CO$_2$)$_{16}$(H$_2$O)$_4$] SMM.[22] Since magnetization measurements were performed on a dried (solvent-free) polycrystalline sample of 1 rather than on wet crystals, it seems incorrect to discuss here relation between the low g tensor value and the some features of crystal and electronic structure of 1.

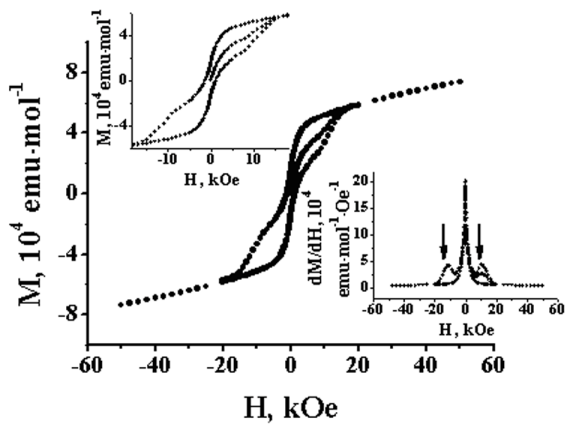

**Fig. 6** Magnetic hysteresis loop at T = 2 K for a dried polycrystalline sample of **1** (coercive force 1200 Oe, the field sweep rate 0.002 T/s). In the insert, the central part of the hysteresis loop (left) and the field dependence of the dM/dH derivative (right) are shown. The steps marked by arrows.

The stepwise magnetization hysteresis loops at low temperatures are also an experimental manifestation of the single-molecule magnetism behavior. The magnetization hysteresis data were obtained at 2K and are shown in Fig. 6. The steps associated with quantum tunneling of the magnetization are not clearly seen on the hysteresis curve. They are discernible on the field dependence of the dM/dH derivative, see the inset in Fig. 6.

**Alternating current (ac) magnetic susceptibility**

To detect the slow relaxation of the magnetization characteristic of the SMM, the ac susceptibility data were collected in the 2-10 K range for six oscillation frequencies from 40 to 1400 Hz. The plots of $\chi_M'T$ versus temperature, where $\chi'$ is the real component of the ac magnetic susceptibility, at various frequencies are shown in Fig. 7.

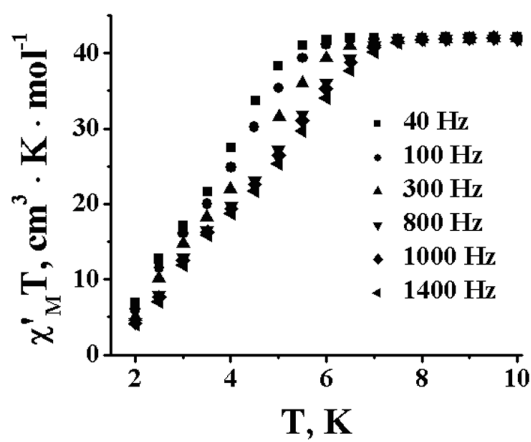

**Fig. 7** Temperature dependence of $\chi'_MT$ for a dried polycrystalline sample of **1**, where $\chi'_M$ is a real component of the molar magnetic susceptibility measured in zero dc field and 4.0 Oe ac field at six frequencies 40, 100, 300, 800, 1000, and 1400 Hz.

A constant value of $\chi_M'T = 42$ cm$^3$K mol$^{-1}$ is observed in the 7-10 K range, which corresponds to the $\chi_M'T$ value expected for a complex with S = 10 (g = 1.74) in the ground state. The values correlate fairly well those obtained from the dc susceptibility measurements. The decrease in $\chi_M'T$ below 6 K is accompanied by the appearance of the systematic frequency dependent out-of-phase $\chi_M''$ ac susceptibility (Fig. 8), the characteristic signature of a superparamagnet-like species such as SMMs[2].

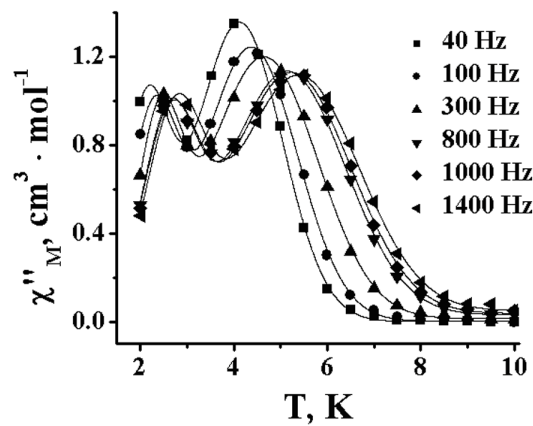

**Fig. 8** Plot of the out-of-phase ac magnetic susceptibility, $\chi''_M$ vs temperature. The data were collected at 4.0 Oe ac field oscillating at six frequencies: 40, 100, 300, 800, 1000, and 1400 Hz.

The plots of $\chi_M''$ versus temperature (Fig. 8) show the peaks in the 4-6 K and the 2-3.5 K ranges indicating the two relaxation processes, which are corresponding to the "high-temperature" (HT) and the "low-temperature" (LT) phases, respectively. These two magnetization relaxation processes in $Mn_{12}$ complexes have been elucidated by a Jahn-Teller isomerism, characteristic of $Mn_{12}$ cluster family.[6,23] The distinction between Jahn-Teller isomers consists in only in the relative orientation of one or more JT axes relative to the $[Mn_{12}O_{12}]$ disk-like core. The HT phases contain Jahn-Teller isomers with a near-parallel alignment of the eight $Mn^{III}$ JT axes along the molecular z-axis (normal orientation, fig. 1b), whereas in the LT phases one or more $Mn^{III}$ JT elongation axis is abnormally oriented equatorially rather than axially. The solvate environment of Mn12 clusters greatly influences the magnetization relaxation processes; in particular, solvent loss causes partial or complete JT isomerisation from the HT phase → LT one or vice versa.[6,20f,23c,d] Since the as susceptibility was measured on a solvent-free sample of complex **1**, one can believe that in our case a loss of the lattice solvent gives rise to a signal from LT phase as a result of Jahn-Teller isomerism.

The $\chi_M''$ vs T plots were used to determine the effective energy barrier ($U_{eff}$) of spin relaxation. Approximation of the resulting relaxation rate ($1/\tau$) versus $T$ dependence by the Arrhenius equation: $\tau=\tau_0\exp(U_{eff}/kT)$, where $\tau$ is the relaxation time, k is the Boltzmann constant, and $\tau_0$ is the pre-exponential term, was done. The relaxation rates at a given temperature can be obtained from $\omega = 2\pi\nu=1/\tau$ at the maxima of the $\chi_M''$ peaks, where $\nu$ is the given oscillation frequency. The plots of $\ln\tau$ vs $1/T$ are depicted in Fig. 9. Approximation of the data results in $U_{eff}$ =58.7 K and $\tau_0$ = 1.3·10$^{-8}$ s for the HT phase and $U_{eff}$ = 34.8 K and $\tau_0$ = 2·10$^{-9}$ s for the LT phase.

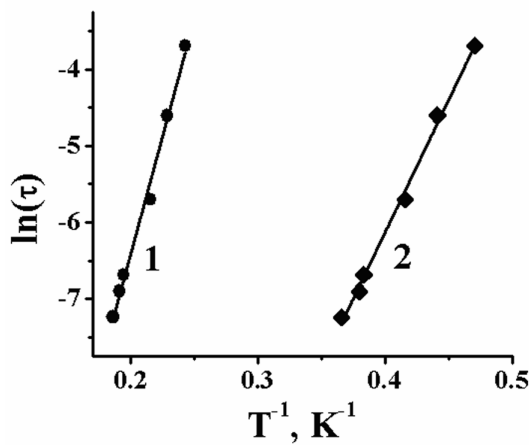

**Fig. 9** Plots of the natural logarithm of the relaxation time, ln(τ), vs inverse temperature for HT (1) and LT (2) phases using $\chi_M''$ vs T data at different frequencies. The solid lines represent least-square fits of the data to the Arrhenius equation.

**Conclusions**

This study has shown that it is possible to only partially substitute the acetate ligands in [Mn$_{12}$O$_{12}$(CH$_3$CO$_2$)$_{16}$(H$_2$O)$_4$] (Mn$_{12}$-acetate) with *o*-cyanobenzoate groups to yield the new single-molecule magnet [Mn$_{12}$O$_{12}$(CN-*o*-C$_6$H$_4$CO$_2$)$_{12}$(CH$_3$CO$_2$)$_4$(H$_2$O)$_4$]·8CH$_2$Cl$_2$ (**1**). In contrast to *o*-cyanobenzoic acid, in the case of *p*-cyanobenzoic acid the complete substitution of acetate ligands takes place.[9c] The distinction between the two isomers of cyanobenzoic acid in the reaction with Mn$_{12}$-acetate is probably associated with the steric effect of *o*-cyanobenzoate ligand containing CN-group located adjacent to carboxylate one. **1** has the high molecular symmetry (S4), which is rather rarely observed among clusters of the Mn$_{12}$ family. From the twelve bridging CN-*o*-C$_6$H$_4$CO$_2^-$ groups in the structure of the complex, eight are axial and four equatorial. The latter are disordered between two orientations. The four other equatorial positions are occupied by the CH$_3$CO$_2^-$ groups. There is, thus, a preference for the CN-*o*-C$_6$H$_4$CO$_2^-$ to occupy the sites lying on the Mn$^{III}$ Jahn-Teller axes. A peculiarity of the structure of **1** is the presence of intermolecular hydrogen bonds C-H···N formed with participation of the axial *o*- cyanobenzoate ligands of the adjacent $Mn_{12}$ clusters. Molecules of the crystallization solvent $CH_2Cl_2$ are arranged into large voids between the $Mn_{12}$ units. They are disordered throughout several sites and are very slightly bonded with the crystal lattice that results in solvent loss immediately after removal of the crystals from the mother liquor.

The calculations of the electronic structure of **1** have shown that, the replacement of acetate ligands with bulky and more electronegative CN-*o*-$C_6H_4CO_2$ groups results in changing of the electronic structure of **1** in comparison to one of the $Mn_{12}$-acetate: *3d* bands of manganese atoms are rather broad, and the center of gravity of the bands shifts down from Fermi level. The overlap between Mn *3d* and O *2p* bands is higher than in $Mn_{12}$-acetate. The changes in electronic structure provide significant difference in the exchange interactions: the values of antiferromagnetic exchanges J(Mn1-Mn2) and J(Mn1-Mn3) decrease, and exchanges between Mn1 atoms in central cubane fragment turn to ferromagnetic and is rather large in **1**, Fig. 3. The main difference of magnetic interactions in **1** compared to other similar $Mn_{12}$ clusters is negligible value of J(Mn1-Mn2).

The direct current magnetization studies on dried polycrystalline sample of **1** in the 2.0-10.0 K and 1.0- 5.0 T ranges were approximated resulting in S =10, $D = -0.56$ K and g = 1.70. The ac susceptibility data exhibits the out-of-phase ($\chi_M''$) signals indicative of slow magnetization relaxation in the 4.0-6.0 K range (HT phase) and 2-3.5 K range (LT phase). The temperature of the $\chi_M''$ peaks is frequency dependent that is characteristic of the SMMs. The availability two phases (HT and LT) is probably due to the Jahn-Teller isomerism, which refers to the orientation of the JT axes of the $Mn^{III}$ in the $Mn_{12}$ core. It is known that the solvate environment of Mn12 clusters greatly affects orientation of the JT axes.[6,23] Since the ac susceptibility was measured on a solvent-free sample of complex **1**, one can believe that in our case the LT phase forms as a result of partial JT isomerisation of the HT phase owing to the loss of lattice solvent under drying.

**Acknowledgements**

The work was financially supported by the Russian Foundation Basic Researches grant №10-03-00128. DWB acknowledges computational support from the CAC of KIAS. The authors thank Dr. S.S. Khasanov and Dr. S.V. Simonov for their contribution to the crystal determination and Prof. R.B. Morgunov for the discussions of the magnetic properties.

**Notes and references**

1. D. Gatteschi, R. Sessoli, *Angew. Chem. Int. Ed.*, 2003, **42**, 268.

2. (a) R.G. Long, *in The Chemistry of Nanostructured Materials*, P. Yang (Ed.), World Sci, Hong Kong, 2003, 1–25; (b) D. Gatteschi, R. Sessoli, J. Villain, *Molecular Nanomagnets*, Oxford University Press, Oxford, UK, 2006; (c) G.E. Kostakis, I.J. Hewitt, A.M. Ako, V. Mereacre, A.K. Powell, *Phil. Trans. R. Soc. A*, 2010, **368**, 1509–1536; (d) X.-Y. Wang, C. Avendano, K.R. Dunbar, *Chem. Soc. Rev.*, 2011, **40**, 3213-3238.

3. (a) M.A. AlDamen, J.M. Clemente-Juan, E. Coronado, C.Marti-Castaldo, A. Gaita-Ariño, *J. Am. Chem. Soc.*, 2008, **130**, 8874-8875; (b) P.-H. Lin, T.J. Burchell, L.Ungur, L..F. Chibotaru, W. Wernsdorfer, M. Murugesu, *Angew. Chem. Int. Ed.*, 2009, **48**, 1-5; (c) R. Sessoli, A.K. Powell, *Coord. Chem. Rev.*, 2009, **253**, 2328-2341; (d) K. Katoh, H. Isshiki, T. Komeda, M. Yamashita, *Coord. Chem. Rev.*, 2011, **255**, 2124-2148; (f) D.-P. Li, T.-W. Wang, C.-H. Li, D.-Sh. Liu, Y.-Z. Li, X.-Z. Yuo, *Chem. Comm.* 2010, **46**, 2929-2931; (e) S.-D. Jiang, B.-W. Wang, H.-L. Sun, Z.-M. Wang, S. Gao, *J. Am. Chem. Soc.*, 2011, **133**, 4730-4733; (h) S.-D. Jiang, B.-W. Wang, G. Su, Z.-M. Wang, S. Gao, *Angew. Chem. Int. Ed.*, 2010, **49**, 7448-7451.

4. (a) N. Magnani, C. Apostolidis, A. Morgenstern, E. Colineau, J.-C. Griveau, H. Bolvin, O. Walter, R. Caciuffo, *Angew. Chem. Int. Ed.*, 2011, **50**, 1696-1698; (b) D.P. Mills, F. Moro, J. McMaster, J. Slageren, W. Lewis, A.J. Blake, S.T. Liddle, *Nature Chemistry*, 2011, **3**, 454-460.

5. (a) A. Caneschi, D. Gatteschi, R. Sessoli, A. L. Barra, L. C. Brunel, M. Guillot, *J. Am. Chem. Soc.*, 1991, **113,** 5873; (b) R. Sessoli, D. Gatteschi, A. Caneschi, M. A. Novak, *Nature*, 1993, **365**, 141-143; (c) R. Sessoli, H.-L. Tsai, A. R. Schake, Sh. Wang, J. B. Vincent, K. Folting, D. Gatteschi, G. Christou, D. N. Hendrickson, *J. Am. Chem. Soc.*, 1993, **115,** 1804-1816.

6. R. Bagai, G. Christou, *Chem. Soc. Rev.*, 2009, **38**, 1011-1026.

7. D. W. Boukhvalov, M. Al-Saqer, E.Z. Kurmaev, A. Moewes, V.R. Galakhov, L. Finkelstein, S. Chiuzbian, M. Neumann, V.V. Dobrovitski, M.I. Katsnelson, A.I. Lichtenstein, B.N. Harmon, K. Endo, J.M. North, N.S. Dalal, *Phys. Rev. B* 2007, **75**, 014419.

8. D. W. Boukhvalov, V. V. Dobrovitski, P. Kogerler, M. Al-Saqer, M. I. Katsnelson, A. I. Lichtenstein, B. N. Harmon, *Inorg. Chem.* 2010, **49**, 10902.

9. (a) T. Kuroda-Sowa, T. Nogami, H. Konaka, M. Maekawa, M. Munakata, H. Miyasaka, M. Yamashita, *Polyhedron*, 2003, **22**, 1795–1801; (b) G.-Q. Bian, T. Kuroda-Sowa, T. Nogami, K. Sugimoto, M. Maekawa, M. Munakata, H.Miyasaka, M. Yamashita, *Bull. Chem. Soc. Jpn.*, 2005, **78**, 1032–1039; (c) V.D. Sasnovskaya, L.A. Kushch, E.B. Yagubskii, I.V. Sulimenkov, V.I. Kozlovskiy, V.S. Zagaynova, T.I. Makarova**,** *J. Magn. and Magn. Mat.***,** 2012, **324**, 2746-2752.

10. W. Wernsdorfer, N. Aliaga-Alcalde, D. N. Hendrickson, G. Christou, *Nature,* 2002, **416**, 406–409.

11. (a) M.N. Leuenberger, D. Loss, *Nature*, 2001, **410**, 789; (b) L. Bogani, W. Wernsdorfer*, Nat. Mater.*, 2008, 7, **179**; (c) T.N. Nguyen, W. Wernsdorfer, K.A. Abboud, G. Christou., *J. Amer. Chem. Soc.*, 2011, **113**, 20688.

12. (a) T. Lis, *Acta Crystallogr. Sect. B,* 1980, **36**, 2042–2046; (b) A. Cornia, R. Sessoli, L. Sorace, D. Gatteschi, A. L. Barra, C. Daiguebonne, *Phys. Rev. Lett.,* 2002, **89**, 257201-25204.

13. M.P. Shores, J.J. Sokol, J.R. Long, *J. Am. Chem. Soc.*, 2002, **124**, 2279.

14. Oxford Diffraction (2007). Oxford Diffraction Ltd., Xcalibur CCD system, CrysAlisPro Software system, Version 1.171.32.

15. G. M. Sheldrick, *Acta Crystallogr., Sect. A*, 2008, **64**, 112.

16. V.I. Anisimov, F. Aryasetiawan, A.I. Lichtenstein, *J. Phys.: Condens. Matter* 1997, **9**, 767–808.

17. U. del Pennio, V. De Renzi, R. Biagi, V. Corradini, L. Zobbi, A. Cornia, D. Gatteschi, F. Bondino, E. Magnano, M. Zangrando, M. Zacchigna, A. Lichtenstein, D.W. Boukhvalov, *Surf. Sci*. 2006, **600**, 4185.

18. O. Gunarson, O. Jepsen, O.K. Andersen, *Phys. Rev. B,* 1983, **27**, 7144.

19. P. Gerbier, D. Ruiz-Molina, N. Domingo, D.B. Amabilino, J. Vidal-Gancedo, J. Tejada, D.N. Hendrickson, J. Veciana, *Monat. Chem.*, 2003, **134**, 265-276.

20. (a) J. An, Z.-D. Chen, X.-X. Zhang, H. G. Raubenheimer, C. Esterhuysen, S. Gao, G.-X. Xu, *J. Chem. Soc., Dalton Trans*., 2000, 3352-3356; (b) H. Zhao, C.P. Berlinguette, J. Basca, A.V. Prosvirin, J.K. Bera, S.E. Tichy, E.J. Schelter, K.R. Dunbar, *Inorg. Chem*., 2004, **43**, 1359-1369; (c) H.-L. Tsai, H.-A. Shia, T.-Y. Jwo, C.-I. Yang, C.-S. Wur, G.-H. Lee, *Polyhedron*, 2005, **24**, 2205-2214; (d) Y.-G. Wei, S.-W. Zhang, M.-C. Shao, Y.-Q. Tang, *Polyhedron*, 1997, **16**, 1471;

(e) L. Zobbi, M. Mannini, M. Pacchioni, G. Chastanet, D. Bonacchi, C. Zanardi, R. Biagi, U. Del Pennino, D. Gatteschi, A. Cornia, R. Sessoli, *Chem. Comm*., 2005, 1640-1642; (f) N.E. Chakov, S.-C. Lee, A.G. Harter, P.L. Kuhns, A.P. Reyes, S.O. Hill, N.S. Dalal, W. Wernsdorfer, K.A. Abboud, G. Christou, *J. Am. Chem. Soc.*, 2006, **128**, 6975-6989;

(g) A-L. Barra, A. Caneschi, A. Cornia, D. Gatteschi, L. Gorini, L-P. Heiniger, R. Sessoli, L. Sorace, *J. Amer. Chem. Soc.,* 2007, **129**, 10754-10762.

21. A. L. Spek, *Acta Crystallogr., Sect A*, 1990, **46**, C34.

22. D. Ruiz-Molina, P. Gerber, E. Rumberger, D. B. Amabilino, I. A. Guzei, K. Folting, J. C. Huffman, A. Rheingold, G. Christou, J. Veciana, D. N. Hendrickson, *J. Mater. Chem*., 2002, **12**, 1152.

23. (a) Z. Sun, D. Ruiz, N.R. Dilley, M. Soler, J. Ribas, K. Folting, m.B. Marle, G. Christou, D.N. Hendrickson, *Chem. Commun*., 1999, 1973-1974; (b) S.M.J. Aubin, Z. Sun, H.J. Eppley, E.M. Rumberger, I.A. Guzei, K. Folting, P.K. Gantzel, A.L. Rheingold, G. Christou, D.N. Hendrickson, *Inorg. Chem*., 2001, **40**, 2127-2146; (c) M. Soler, W. Wernsdorfer, Z. Sun, J.C. Huffman, D.N. Hendrickson, G. Christou, *Chem. Comm*., 2003, 2672-2673; (d) A.J. Tasiopoulos, W. Wernsdorfer, K.A. Abboud, G. Christou, *Inorg. Chem.*, 2005, **44**, 6324-6338.